\documentclass{ws-procs9x6}
\usepackage{graphicx}
\usepackage{amsmath}
\usepackage{url}

\newcommand{\compact}[1]{}    

\newcommand{\ForArXiv}[1]{#1}

\newcommand{\as}{\alpha_s}
\newcommand{\cN}{{\mathcal N}}

\newcommand{\ie}{\emph{i.e.}\ }
\newcommand{\eg}{\emph{e.g.}\ }
\newcommand{\order}[1]{\mathcal{O}\left(#1\right)}

\newcommand{\ee}{e^+e^-}
\newcommand{\jet}{\,\mathrm{jet}}
\newcommand{\jets}{\,\mathrm{jets}}

\begin{document}

\title{RECENT DEVELOPMENTS IN PERTURBATIVE QCD%
  \ForArXiv{\footnote{\uppercase{I}nvited plenary talk at the \uppercase{XIV
      I}nternational \uppercase{W}orkshop on \uppercase{D}eep
    \uppercase{I}nelastic \uppercase{S}cattering (\uppercase{DIS}
    2006), \uppercase{A}pril 2006, \uppercase{T}sukuba,
    \uppercase{J}apan.}}  
}

\author{ Gavin P. SALAM }

\address{LPTHE, CNRS UMR 7589;
  Universit\'e P. et M. Curie (Paris VI); Universit\'e
  Denis Diderot (Paris VII),
  75252 Paris cedex 05, France}

\maketitle\abstracts{ A brief overview of some recently active topics
  in perturbative QCD, including: string-inspired recursion techniques
  at tree level; recursion approaches and automation of standard
  techniques for 1-loop calculations; the status of NNLO jet
  calculations; and non-trivial structures that appear in higher-order
  calculations.  }

\ForArXiv{
\vspace{-8.5cm}
\begin{flushright}
  LPTHE-P06-05\\
  hep-ph/0607153 \\
  July 2006
\end{flushright}
\vspace{5.8cm}}

\section{Introduction}

As the startup of LHC approaches, much current work in QCD is directed
towards developing techniques for improving the flexibility and
accuracy of perturbative calculations.

Flexibility (section~\ref{sec:multi-jet}) is crucial because of the
vast range of multi-jet final states that will be studied in LHC
new-particle searches. At tree level, numerical recursion techniques
have long been used to build multi-leg amplitudes from
amplitudes with fewer legs --- recent developments inspired by
string theory have led to analytically more powerful recursions,
giving many new compact results for tree-level amplitudes. This new
understanding is also being applied to 1-loop amplitudes, often the
missing ingredient for quantitatively reliable (NLO) multi-jet
predictions. In parallel, more traditional 1-loop techniques are being
subjected to automation, and here too major progress has recently been
made.

Accuracy (section~\ref{sec:prec-QCD}), in the sense of NNLO jet
calculations, 
is looking like it might be %
within reach in the coming year. This is welcome since for over a
decade LEP and HERA have been delivering final-state measurements with
precisions several times better than the NLO theory uncertainties, and
the latter limit our ability to extract fundamental parameters of QCD
such as $\as$ and parton distributions.  NNLO results also provide
clues as to the general structures of high orders in QCD. This is both
of fundamental interest and potentially useful in predicting large
parts of yet higher orders.

\section{Multi-jets}
\label{sec:multi-jet}

\subsection{Tree level}
\label{sec:tree-level}

\begin{figure}[b]
  \centering
  \includegraphics[width=0.85\textwidth]{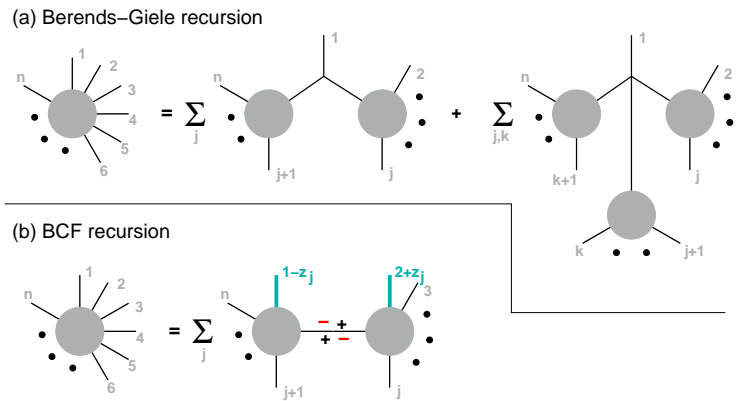}\vspace{-1em}
  \caption{Graphical comparison of
    Berends-Giele\protect\cite{BerendsGiele} and BCF\protect\cite{BCF}
    recursion relations.}
  \label{fig:BerGie+BCF}
\end{figure}

The multiplicity of Feynman graphs grows factorially with the number
of legs of a process, hampering the usefulness of traditional
techniques for calculating multi-leg amplitudes. 
An important discovery of the 1980's was the Berends-Giele
recursion,\cite{BerendsGiele} allowing amplitudes to be constructed by
assembling smaller, off-shell sub-amplitudes,
fig.\ref{fig:BerGie+BCF}a. This was suited to recursive numerical
evaluations and helped prove analytical all-order
results,\cite{ParkeTaylor} thanks to simplifications that occur
independently in each sub-amplitude. Berends-Giele recursion joins
amplitudes via three and four-gluon vertices. Recently two new
recursions were discovered, CSW\cite{CSW} and BCF\cite{BCF}
(fig.\ref{fig:BerGie+BCF}b) which join amplitudes via a scalar
propagator. In the latter case the amplitudes are made on-shell using
analytic continuation of reference momenta (legs 1,2 in figure). The
simpler structure of CSW and BCF recursion makes it easier to identify
simplifications at each order, leading to many new all-order
analytical results, including applications to processes with heavy
quarks and electroweak bosons (for a review, see
ref.~\refcite{DixonReview}). Though originally inspired from string
theory, the new recursion relations have been proved based on general
field theoretical arguments\cite{BCF,Risager} (exploiting the
rationality of tree-level amplitudes) and can also be related directly
to Feynman diagrams.\cite{FeynmanBCFW}

The above developments represent an impressive improvement in our
understanding of field theory. Nevertheless one should bear in mind
that for practical numerical implementations of tree level
calculations, existing methods still remain competitive.\cite{Worek}

\subsection{NLO}
\label{sec:NLO}

While many NLO calculations already exist,\cite{hepcode} there is a
recognized phenomenological need for further multi-leg NLO
calculations, in particular to simulate backgrounds to new physics
signals (for a full discussion of recent developments, see
ref.\refcite{LesHouches}).
%
%
The ingredients in
$n$-leg NLO calculations are $n+1$ leg tree amplitudes, $n$-leg 1-loop
amplitudes and a procedure for combining the pieces. The
hardest part is the $1$-loop calculation, for which several $5$-leg
results exist and some first $6$-leg results are starting to appear.

The string theory inspired approach to tree-level diagrams does not
trivially generalise to loop level, in part because of the more
complicated analytical structure of loop diagrams (cuts as well as
poles). Nevertheless the string-inspired approach has led to much new
work on loops as well, notably using the ``sewing together'' of tree
diagrams. This works most easily for supersymmetric loop amplitudes,
where cancellations between scalar, fermionic and vector particles in
the loop lead to simpler structures in the final answer.  The plain
QCD result is then obtained by combining answers with $\cN=4$ SUSY,
$\cN=1$ SUSY and a scalar particle in the loop, the latter being the
most difficult (split into ``cut-constructable'' ($c,d,e$) and purely
rational ($R$) pieces). Considerable progress has been made, as
illustrated in table~1 which shows the contributions to the analytical
evaluation of the six-gluon 1-loop amplitude, for all independent
helicity configurations. For MHV configurations (two $-$ helicities,
all others $+$) and split NMHV and NNMHV configurations (3 or 4
adjacent $-$ helicities, all others $+$), general multiplicity results
have very recently become available.\cite{BBDFK06,BBDFK06b}

\begin{table}[b]
  \centering
  \tbl{The analytically derived helicity components of the 1-loop 6-gluon
    amplitude (adapted from ref.~\protect\refcite{TwistorLoopReview}). }
  {\small
  \begin{tabular}{|l|l|l|l|l|} \hline
    & $\cN=4$    & $\cN=1$     
    & $ S\, (c,d,e) $  &   $ S\, (R) $
    \\[2pt]
    \hline
    {\small  $A(--++++)$ }  &  [\refcite{BDDK94a}]     &
    [\refcite{BDDK94b}]    & [\refcite{BDDK94b}]    &  [\refcite{BDK05}] 
    \\[2pt]
    \hline
    {\small $A(-+-+++) $ }  & [\refcite{BDDK94a}]     &
    [\refcite{BDDK94b}] & [\refcite{BBST04}]  & [\refcite{BBDFK06b,XYZ06}] 
    \\[2pt]
    \hline
    {\small $A(-++-++)$ }  & [\refcite{BDDK94a}]     &
    [\refcite{BDDK94b}]   & [\refcite{BBST04}] & [\refcite{BBDFK06b,XYZ06}]   
    \\[2pt]
    \hline
    {\small $A(---+++)$ }  &  [\refcite{BDDK94b}]   &
    [\refcite{BBDD04}]  & [\refcite{BBDI05}]  & [\refcite{BBDFK06}]  
    \\[2pt]
    \hline
    {\small $A(--+-++)$}  &  [\refcite{BDDK94b}]    &
    [\refcite{BBCF05,BBDP04+5}]    & [\refcite{BFM06}] &  [\refcite{XYZ06}]
    \\[2pt]
    \hline
    {\small $A(-+-+-+)$}  &   [\refcite{BDDK94b}]   &
    [\refcite{BBCF05,BBDP04+5}]    & [\refcite{BFM06}]  &  [\refcite{XYZ06}]
    \\[2pt]
    \hline
  \end{tabular}}
\caption{}
\label{tab:loopstatus}
\end{table}

An alternative approach automates traditional methods, \ie Feynman
diagram generation, and the recursive reduction of the resulting loop
integrals to a set of known basis integrals. It has the advantage of
being easier to generalise to processes with external particles other
than gluons, but  suffers from the large number of Feynman
diagrams, each term of which is broken up into many further terms by
the (sometimes numerical) recursion.  Sometimes the recursion
introduces numerical instabilities and alternative strategies are then
required.\cite{Binoth:2005ff,Ellis:2005zh} A notable result with such
methods was the first full evaluation of the $6$-gluon 1-loop
amplitude for arbitrary helicity configurations,\cite{EGG06-6gluon}
and work is in progress for the $6$-quark 1-loop
amplitude.\cite{Binoth:2006mf} Full $2\to4$ NLO jet predictions are
however still some way off.

Related automated methods have been successful also in electroweak
calculations, with recent full results for $e^+e^- \to
4$~fermions\cite{DennerDittmaier} and $e^+e^- \to
HH\nu\bar\nu$,\cite{Boudjema:2005rk} and progress made there will
hopefully in part carry over to QCD.  Also, traditional techniques can
simplify considerably\cite{XYZ06} when extracting just the
scalar rational components of the decomposition in table~1 (\ie the
parts hardest to obtain in the string-related approaches).

\section{Precision QCD}
\label{sec:prec-QCD}

\subsection{NNLO jets}
\label{sec:NNLO-jets}

Various results exist at NNLO for processes with two QCD partons at
Born level and one or two non-QCD particles. The current challenge is
to address processes with three or more QCD legs at Born level, in
particular $\ee \to 3\jets$. All tree-level, 1 and 2-loop amplitudes
are known --- the difficulty is in cancelling divergences between them
for a general jet observable.

Two approaches exist. Subtraction (as at NLO\cite{dipole}) identifies
a function with the same divergences as the real amplitudes, but that
is sufficiently simple that it can be integrated analytically --- one
then subtracts the unintegrated form from the real amplitudes and adds
the integrated form to the virtual amplitudes, cancelling all
divergences. Finding the subtraction functions requires deep
understanding of the QCD divergences and ingenuity so as to make the
result integrable. A full scheme at NNLO for processes with just
final-state particles has been proposed\cite{G3NLO} and as a proof of
concept used to calculate to the $\as^3/N_c^3$
contribution to the mean thrust in $e^+e^-$. 

An alternative approach, sector decomposition,\cite{SectorDecomp}
rewrites phase space to as to isolate single divergences and then
effectively introduces plus-prescriptions (as in splitting functions)
so as to allow separate extraction of different powers of the
dimension regularisation $\epsilon$. 
This is less dependent on the specific structure of QCD divergences,
but becomes more complicated as the number of QCD particles increases.
It has been successfully used for hadron-hadron processes with two
Born QCD particles,\cite{Anastasiou:2003ds} and for a part of the NNLO
$\ee\to3\jets$ cross section.\cite{Heinrich:2006sw}

Given the above progress one can perhaps expect first full NNLO
predictions for $\ee\to3\jets$ in the coming year, hopefully with a
major impact on measurements of the coupling and studies of analytical
hadronisation models. Extensions to DIS $2+1\jet$ events and
hadron-collider dijets will probably take somewhat longer.  Note that
for jets at hadron colliders, an issue remains with the experimental
jet definitions. Because the standard midpoint cone
(ILCA\cite{RunIIJets}) has the drawback that it can leave large energy
deposits unclustered,\cite{Ellis:2001aa} an extra `search-cone' step
that has been proposed\cite{Ellis:2001aa} and used.\cite{CDFcone}
However this turns out to be infrared (IR) unsafe as the seed
threshold is taken to zero,\cite{Wobisch} 
compromising theory-data comparisons.  A positive development is that
hadron-collider measurements with the more physically motivated (and
IR safe) $k_t$ algorithm have been shown to be feasible now by both
Tevatron collaborations,\cite{D0kt,CDFkt} and the long-standing speed
issue for the $k_t$ algorithm at high-multiplicity has also been
resolved.\cite{FastJet}

\subsection{Structure of perturbation theory}
\label{sec:struct-pt}

Two years have passed since Moch, Vermaseren and Vogt's (MVV) seminal
calculation of the NNLO splitting functions.\cite{Moch:2004pa}
With related technology, the same authors have obtained the third
order coefficient functions,\cite{3rdOrderCoeff} threshold resummation
coefficients,\cite{MVVThreshold} and quark and gluon form
factors.\cite{QGFormFactors} These results have served as
ingredients to calculations of 3-loop $\cN=4$ SUSY splitting
functions,\cite{Kotikov:2004er} Drell-Yan and Higgs threshold
resummations,\cite{ThresholdDYH} and 3-loop non-singlet time-like
splitting functions.\cite{Mitov:2006ic}

Various unexpected structures appear in the above results. 
E.g.\ 
writing
\begin{equation}
  \label{eq:Pstruct}
  P_{ij}(x) = \frac{A}{(1-x)_+} + B\delta(1-x) + C\ln(1-x) + \order{1}
\end{equation}
with $A = \sum_n A_n (\as/4\pi)^n$, etc., it was noted at
NLO\cite{Curci:1980uw} that $C_2 = A_1^2$. At NNLO, MVV observed $C_3
= 2A_1A_2$. If one postulates splitting functions to be
universal\cite{Dokshitzer:1995ev} (identical for time and space-like
evolution) when expressed for a modified evolution variable $z^\sigma
Q^2$ ($\sigma = \pm 1$ for the $^\mathrm{time}_\mathrm{space}$-like
case)\footnote{Specifically $\partial_{\ln Q^2} D = \int \frac{dz}{z}
  {\mathcal P}(z,Q^2) D(\frac{x}{z},z^\sigma Q^2)$, with $D$ a parton
  distribution or fragmentation function.} %
and furthermore assumes the universal splitting function to be
classical at large $x$ (having $C\equiv0$), then for normal space-like
splitting functions one predicts that $C = A^2$ at all
orders,\cite{Dokshitzer:2005bf} precisely as found at NLO and NNLO.

The idea of a universal splitting function is given further
credibility by an analysis\cite{Mitov:2006ic} which uses the
usual\cite{Curci:1980uw} analytical continuation $x \to 1/x$ to go
from the space-like to the time-like non-singlet (NS) case and finds
it to be identical to the time-like result found assuming
universality with the $z^\sigma Q^2$ evolution variable. 
Note that
universality predicts the 3-loop
$P_{NS}^{\sigma=+1} - P_{NS}^{\sigma=-1}$ difference using only 
2-loop information. Given that the full 3-loop $P_{NS}^{\sigma=+1}$
and $P_{NS}^{\sigma=-1}$ are themselves also related by $x \to 1/x$,
this implies the existence of  non-trivial (and yet to be
understood) properties of the analytic structure of the splitting
functions. The universality also suggests an explanation for the
till-now mysterious absence of 2 and 3-loop leading log $x$ terms in
the space-like splitting functions, as being closely related to exact
angular order in fragmentation.\cite{Marchesini:2006ax} Despite these
successes the universality hypothesis requires further development
notably as concerns the treatment of the singlet sector and the
factorisation scheme.

Other intriguing perturbative structures that have also been found
recently include the following: in $\cN=4$ SUSY QCD there is
increasing evidence that $n$-loop $m$-leg amplitudes are
related to the $n^\mathrm{th}$ power of the $1$-loop $m$-leg
amplitude\cite{N4SusyLoops} (new numerical methods\cite{NumLoops} for
loop calculations providing powerful cross checks); in large-angle
soft-gluon resummation for $2\to 2$ scattering, there is a mysterious
symmetry\cite{Dokshitzer:2005ig} when exchanging the kinematic
quantity $(\ln s^2/ut - 2\pi)/(\ln u/t)$ and the number of colours,
$N_c$.

\section{Other results}

Owing to limitations of space, many active topics have been omitted.
Some (small-$x$ saturation, generalised parton distributions) are
reviewed in these proceedings.\cite{DiehlIancuTheseProc} A more
extensive bibliography is to be found in ref.~\refcite{Salam:2005du}.
For others new developments, the reader is referred to the literature,
notably for 4-loop decoupling relations for
$\as$;\cite{4-loop-decouple} jet definitions that preserve the IR
safety of flavour;\cite{Banfi:2006hf} the release of the first
\texttt{C++} \texttt{ThePEG}-based hadron-collider Monte Carlo (MC)
generator;\cite{Gieseke:2006rr} progress in practical and conceptual
aspects of matching MC and NLO;\cite{MC@NLO} reweighting to match MC
with NNLL and NNLO;\cite{reweighting} and soft large-angle
resummations, both in terms of phenomenology,\cite{Forshaw:2005sx}
understanding of treatment of jet-algorithms for non-global
resummation,\cite{Banfi:2005gj} two-loop soft colour evolution
matrices\cite{Aybat:2006wq} and other recent NNLO resummation
results,\cite{NNLOResum} and an intriguing (but still to be confirmed)
suggestion of a breakdown of coherence at high
orders.\cite{Forshaw:2006fk}

\section*{Acknowledgments}
I am grateful to the many colleagues who provided me with insight over
the past year on the topics discussed here.
%
%
I  thank also the organisers of DIS 2006 for their kind
invitation (and financial support) to this active and stimulating
workshop.

\end{document}